\documentclass[sn-aps]{sn-jnl}

\usepackage{graphicx}%
\usepackage{multirow}%
\usepackage{amsmath,amssymb,amsfonts}%
\usepackage{amsthm}%
\usepackage{mathrsfs}%
\usepackage[title]{appendix}%
\usepackage{xcolor}%
\usepackage{textcomp}%
\usepackage{manyfoot}%
\usepackage{booktabs}%
\usepackage{bm}
\usepackage{listings}%

\begin{document}

\title[Momentum dependent nucleon-nucleon contact interactions and
their effect on $p-d$ scattering observables]{Momentum dependent nucleon-nucleon contact interactions and
their effect on $p-d$ scattering observables}


\date{\today}
\author*[1,2]{\fnm{E.} \sur{Filandri}}\email{elena.filandri@df.unipi.it}

\author[3,4]{\fnm{L.} \sur{Girlanda}}
\author[1,2]{\fnm{A.} \sur{Kievsky}}
\author[1,2]{\fnm{L.E.} \sur{Marcucci}}
\author[1,2]{\fnm{M.} \sur{Viviani}}

\affil*[1]{\orgdiv{Department of Physics}, \orgname{University of Pisa}, \orgaddress{\street{Largo Bruno Pontecorvo}, \city{Pisa}, \postcode{56127},  \country{Italy}}}

\affil[2]{ \orgname{INFN Sezione Pisa}, \orgaddress{\street{ Largo Bruno Pontecorvo, 3/Edificio C}, \city{Pisa}, \postcode{56127}, \country{Italy}}}

\affil[3]{\orgdiv{Department of Mathematics and Physics}, \orgname{University of Salento}, \orgaddress{\street{Via per Arnesano}, \city{Lecce}, \postcode{I-73100}, \country{Italy}}}

\affil[2]{\orgname{INFN, Sez. di Lecce},  \orgaddress{ \city{Lecce},\postcode{I-73100}, \country{Italy}}}

\abstract{Starting from a complete set of relativistic nucleon-nucleon contact operators  up to order $O(p^4)$ of the expansion  in the soft (relative or nucleon) momentum  $p$, we show that non-relativistic expansions of relativistic operators involve twenty-six independent combinations, two starting at $O(p^0)$, seven at order $O(p^2)$ and seventeen at order $O(p^4)$. This demonstrates the existence of two low-energy free constants that parameterize  interactions dependent on the total momentum of the pair of nucleons $P$.
The latter, through the use of a unitary transformation, can be removed  in the two-nucleon  fourth-order contact interaction  of the Chiral Effective Field Theory, generating a three-nucleon  interaction at the same order.
Within a hybrid approach in which this interaction is considered together with the phenomenological potential AV18, we show that the LECs involved can be used to fit very accurate data on the polarization observables of the low-energy $p-d$ scattering, in particular the $A_y$ asymmetry. }




\maketitle

\section{Introduction}
Effective Field Theories (EFTs)\cite{Weinberg:1990rz,Weinberg:1991um,Weinberg:1992yk,Bedaque:2002mn,Epelbaum:2005pn,Epelbaum:2008ga,Machleidt:2011zz} have established themselves as the preferred systematic approach for tackling the complex problem of nuclear interactions. This approach rests on several fundamental principles. It starts with the identification of the most general effective Lagrangian, respecting all pertinent symmetries, including the approximate chiral symmetry of Quantum Chromodynamics (QCD). The ordering of interactions is accomplished through a power-counting scheme. Consequently, this framework yields a predictive context in which physical observables, at each stage of the low-energy expansion, are expressed in terms of a finite set of low-energy constants (LECs). These LECs serve as fitting parameters and are determined through experimental data.

One essential task is the precise determination of the required number of parameters, both necessary and sufficient, at each stage of the expansion. This task not only rigorously scrutinizes the theory but also aids in estimating the theoretical uncertainty arising from unaccounted higher-order interactions \cite{Furnstahl:2014xsa,Epelbaum_2015,Furnstahl:2015rha,Konig:2019adq}. In the realm of nuclear forces, these fitting parameters pertain to LECs associated with contact interactions between nucleons, which are not constrained by chiral symmetry. However, they are subject to constraints imposed by Poincaré symmetry \cite{Foldy:1960nb,Krajcik:1974nv}. Despite the common non-relativistic quantum-mechanical framework used in nuclear physics,  Poincaré symmetry must ultimately be respected. This requires the reconciliation of various relativistic effects arising from different sources, such as recoil corrections in energy denominators and vertex corrections from the heavy baryon expansion.
Given that relativistic effects scale with the soft nucleon momenta, they can be systematically examined in the low-energy expansion, and the constraints on interactions can be imposed algebraically.

In the present work, starting from a manifestly Lorentz-invariant two-nucleon ($2N$) contact Lagrangian density and performing non-relativistic reductions up to the order $1/m^4$, where $m$ represents the mass of the nucleon we retrace the results already obtained in Refs. \cite{Xiao:2018jot,Filandri_2023}.

Furthermore, we  underline the importance of  two additional $2N$ contact LECs at N3LO, which characterize momentum-dependent interactions allowed by Poincaré symmetry. These LECs can be transformed into a three-nucleon ($3N$) interaction through a unitary transformation. This finding may explain the challenges faced when attempting to enhance accuracy in $3N$ systems, particularly in the context of scattering observables, during the transition from N2LO to N3LO \cite{Golak}.

The inclusion of the N4LO $3N$ contact interaction has proven to be of significant importance in reducing existing discrepancies between theoretical predictions and experimental data \cite{Girlanda:2023znc}.

This paper offers quantitative evidence that the additional two LECs $D_{16}$ and $D_{17}$ at N3LO introduce the necessary flexibility to substantially enhance the description of low-energy $p-d$ scattering polarization observables, with a particular focus on the $A_y$ asymmetry. This aspect has long posed challenges for most nuclear interaction models.

Our approach is hybrid in nature, involving the consideration of the $3N$ force induced  by  $D_{16}$ and $D_{17}$ potential terms in conjunction with the phenomenological AV18 $2N$ potential. A more comprehensive calculation in Chiral Effective Field Theory (ChEFT) is deferred to future research.

The structure of the paper is as follows.
In Sec. \ref{sec1} we show the basic steps of non-relativistic reduction to order $p^4$ of a covariant $2N$ Lagrangian, emphasizing the existence of two interactions dependent by the total momentum $P$  accompanied by two free LECs $D_{16}$ and $D_{17}$.
In Sec. \ref{sec2} we explain how these two off-shell interactions are related by unitary transformation to a three-body force and how they can be used for a fit of $p-d$ polarization observables. In Sec. \ref{sec3} we show the fit results and in Sec. \ref{sec4} final conclusions are drawn.

\section{Two extra interactions from the non relativistic reduction of $2N$ Contact Lagrangian up to N3LO}\label{sec1}
The general expression of the relativistic $2N$ contact Lagrangian is derived following the approach of Ref.~\cite{Xiao:2018jot,Filandri_2023,Girlanda:2010zz, Petschauer:2013uua}. It consists of products of fermion bilinears, such as
\begin{equation}
  (\bar{\psi}\overleftrightarrow{\partial}_{\mu_1}\cdots\overleftrightarrow{\partial}_{\mu_i}\Gamma_A\psi)\partial_{\lambda_1}\cdots\partial_{\lambda_k}(\bar{\psi}\overleftrightarrow{\partial}_{\nu_1}\cdots\overleftrightarrow{\partial}_{\nu_j}\Gamma_B\psi),
\end{equation}
where $\psi$ indicates the relativistic nucleon field, a doublet in isospin space, and $\Gamma_{A,B}$ are generic elements of the Clifford algebra.

To construct the covariant Lagrangian, various symmetries must be satisfied, including Lorentz invariance, isospin, parity, charge conjugation, and time reversal symmetry. According to the CPT theorem, time reversal symmetry is automatically satisfied if charge conjugation and parity symmetries are fulfilled.

Table \ref{tab:Trans.Pr} lists the transformation properties of different elements of the Clifford algebra, metric tensor, Levi-Civita tensor, and derivative operators under parity ($\mathcal{P}$), charge conjugation ($\mathcal{C}$), and Hermitian conjugation (h.c.). These properties play a crucial role in the construction of the Lagrangian.
 \begin{table}
\label{tab:Trans.Pr}
\caption{Transformation proprieties of the different elements of the Clifford algebra, metric tensor, Levi-Civita tensor and derivative operators under parity ($\mathcal{P}$), charge conjugation ($\mathcal{C}$) and Hermitian conjugation (h.c.)}
\centering
 \begin{tabular}{|p{0.1cm} p{0.1cm} p{0.1cm} p{0.1cm} p{0.4cm} p{0.2cm} p{0.2cm} p{0.3cm} p{0.2cm} p{0.1cm} p{1cm}|}
\hline
    &$1$&$\gamma_5$&$\gamma_\mu$&$\gamma_\mu\gamma_5$&$\sigma_{\mu\nu}$&$g_{\mu\nu}$&$\epsilon_{\mu\nu\rho\sigma}$&$\overleftrightarrow{\partial}_\mu$&$\partial_\mu$& $\tau^{a}$\\
\hline
$\mathcal{P}$   &$+$&$-$&$+$&$-$&$+$&$+$&$-$&$+$&$+$ & $+$\\
$\mathcal{C}$   &$+$&$+$&$-$&$+$&$-$&$+$&$+$&$-$&$+$ & $(-1)^{a+1}$\\
h.c.    &$+$&$-$&$+$&$+$&$+$&$+$&$+$&$-$&$+$& $+$\\
\hline
\end{tabular}
\end{table}

Regarding the isospin degrees of freedom, both flavor structures $1\otimes1$ and $\tau^a\otimes\tau^a$ are allowed, but the latter can be disregarded thanks to Fierz identities. To specify the chiral order of each building block, it is necessary to identify the powers of the soft relative momentum ${\bf p}$.
The derivatives $\partial$ acting on the entire bilinear are of order $p$, while the derivative $\overleftrightarrow{\partial}$ acting inside a bilinear is of order  $p^0$ due to the presence of the heavy fermion mass scale. The fields equations of motion can be used to reduce the number of cosidered terms.
Further criteria in specifying the power counting of the operators regard the Dirac matrix $\gamma_5$, which can be thought of as $O(p)$ since it mixes the large and small components of the Dirac spinor, and the Levi-Civita tensor $\epsilon_{\mu\nu\rho\sigma}$, which raises the chiral order by $n-1$, when contracted with $n$ derivatives acting inside a bilinear.

These guidelines lead to the complete set of relativistic contact operators displayed in Table 2 of Ref.\cite{Filandri_2023}. The last column contains, for each one of the Dirac structures, the additional combination of four-gradients arising up to $O(p^4)$.
This construction differs from the one conducted in Ref.~\cite{Xiao:2018jot}, due to a different choice of operators reduced by the equations of motion. 

The next step is the non-relativistic reduction of these operators in terms of a minimal basis of non-relativistic $2N$ contact operators, involving up to 4 powers of three-gradients in terms of non-relativistic nucleon fields. It is important to note that the $2N$ contact Hamiltonian density takes the form of ${\cal H}_{2N} = {\cal H}^{(0)} + {\cal H}^{(2)} + {\cal H}^{(4)}$, with ${\cal H}^{(0)}$, ${\cal H}^{(2)}$, and ${\cal H}^{(4)}$ defined with the corresponding LECs.

Table 3 of Ref.\cite{Filandri_2023} provides a complete basis of non-relativistic operators computed between states of two nucleons with initial and final momenta. It includes both LECs and operators for $O(p^4)$. Notably, this basis accounts for the general reference frame, and indeed some of the operators are $P$-dependent, where $P$ denotes the overall momentum of the nucleon pairs.

The operators related to the constants $D_{16}$ and $D_{17}$ are introduced in this basis, representing new LECs that parametrize the  $P$-dependent $2N$ interaction in the general reference frame. These LECs do not contribute in the center-of-mass frame.

\section{Influence of two-body off-shell forces on the $A_y$ puzzle}\label{sec2}

The N3LO $2N$ contact potential was originally considered in Refs.~\cite{machleidt_plb02,machleidt_prc03} as consisting of 15 LECs. After careful scrutiny of the constraints imposed by relativity, two further LECs emerge, leading to the following expression in the general reference frame,  
 \begin{align}
 V_{2N}^{(4)}=& D_{1} k^{4}+D_{2} Q^{4}+D_{3} k^{2} Q^{2}+D_{4}(\bm{k} \times \bm{Q})^{2} +\left[D_{5} k^{4} \right.\nonumber \\&\left. +D_{6} Q^{4}+D_{7} k^{2} Q^{2}+D_{8}(\bm{k} \times \bm{Q})^{2}\right]\left(\bm{\sigma}_{1} \cdot \bm{\sigma}_{2}\right) \nonumber\\
&+\frac{i}{2}\left(D_{9} k^{2}+D_{10} Q^{2}\right)\left(\bm{\sigma}_{1}+\bm{\sigma}_{2}\right) \cdot(\bm{Q} \times \bm{k})\nonumber\\&+\left(D_{11} k^{2}+D_{12} Q^{2}\right)\left(\bm{\sigma}_{1} \cdot \bm{k}\right)\left(\bm{\sigma}_{2} \cdot \bm{k}\right)\nonumber\\&+\left(D_{13} k^{2}+D_{14} Q^{2}\right)\left(\bm{\sigma}_{1} \cdot \bm{Q}\right)\left(\bm{\sigma}_{2} \cdot \bm{Q}\right) \nonumber\\&
+D_{15} \,\bm{\sigma}_{1} \cdot(\bm{k} \times \bm{Q}) \,\bm{\sigma}_{2} \cdot(\bm{k} \times \bm{Q})\nonumber\\&+i D_{16}\, \bm{k} \cdot \bm{Q}\, \bm{Q} \times \bm{P} \cdot\left(\bm{\sigma_{1}}-\bm{\sigma_{2}}\right)\nonumber\\&+D_{17}\, \bm{k} \cdot \bm{Q}\,(\bm{k} \times \bm{P}) \cdot\left(\bm{\sigma_{1}} \times \bm{\sigma_{2}}\right)\, \label{eq:2N}
\end{align}
with $\bm{ k}=\bm{ p}'-\bm{ p}$ and $\bm{ Q}=\frac{\bm{ p}'+\bm{ p}}{2}$, $\bm{ p}$ and $\bm{ p}'$ being the initial and final relative momenta, and $\bm{ P}=\bm{p}_1+\bm{ p}_2$ the total pair momentum.
However, as it was pointed out in Ref.~\cite{Reinert},  only 12 independent LECs survive on shell and can thus be determined from $2N$ scattering data.
This redundancy amounts to a unitary ambiguity, i.e. to the possibility of generating shifts of the LECs by unitary transforming the one-body kinetic energy operator $H_0$ as
\begin{equation}
    H_0 \to  U^\dagger H_0  U.
\end{equation}
Here $U$ is the most general unitary 2-body contact transformation depending on 5 arbitrary parameters $\alpha_i$,
\begin{equation} \label{eq:unitaryh0}
    U = \exp \left[\sum_{i=1}^5 \alpha_i T_i\right],
\end{equation}
and the independent generators $T_i$, which are  given explicitly in Ref.~\cite{Girlanda_2020,Girlanda:2023znc}, induce a shift in the N3LO contact LECs, $D_i \to D_i + \delta D_i$. Specifically, the induced shifts for $D_{16}$ and  $D_{17}$ are given by:
\begin{align}
\delta D_{16} &= -\frac{2}{m}\alpha_4, \label{eq:deltaD16} \\
\delta D_{17} &= -\frac{4}{m}\alpha_3 - \frac{2}{m}\alpha_5. \label{eq:deltaD17}
\end{align}

At the same time, the unitary transformation, when applied to the LO $2N$ contact Hamiltonian, induces a shift of the  LECs parametrizing the subleading three-body interaction $E_i$, as
\begin{eqnarray} \label{eq:induced3n}
V^{(2)}_{3N,\Lambda}&=&\sum_{i j k}  \left[  E_1 + E_2 {\bm \tau}_i \cdot {\bm \tau}_j +  \left(  E_3  +  E_4 {\bm \tau}_i \cdot {\bm \tau}_j \right)  {\bm \sigma}_i \cdot {\bm \sigma}_j\right] \nonumber \\
&& \times \left[ Z_\Lambda^{\prime\prime}(r_{ij}) + 2 \frac{Z_\Lambda^\prime(r_{ij})}{r_{ij}}\right] Z_\Lambda(r_{ik})  \nonumber \\
&& + (  E_5 + E_6 {\bm \tau}_i\cdot{\bm \tau}_j) S_{ij} \left[ Z_\Lambda^{\prime\prime}(r_{ij}) - \frac{Z_\Lambda^\prime(r_{ij})}{r_{ij}}\right] Z_\Lambda(r_{ik}) \nonumber \\
&&+ ( E_7 + E_8 {\bm \tau}_i\cdot{\bm \tau}_k) \left\{ ( {\bf L}\cdot {\bm S})_{ij} ,\frac{Z_\Lambda^\prime(r_{ij})}{r_{ij}} Z_\Lambda(r_{ik}) \right\} \nonumber \\
&& +\left[  (E_9 +  E_{10} {\bm \tau}_j \cdot {\bm \tau}_k) {\bm \sigma}_j \cdot \hat{{\bm r}}_{ij}  {\bm \sigma}_k \cdot \hat{ {\bm r}}_{ik}   \right. \nonumber \\
&&+\left. ( E_{11} +  E_{12} {\bm \tau}_j \cdot {\bm \tau}_k+  E_{13} {\bm \tau}_i \cdot {\bm \tau}_j) {\bm \sigma}_k \cdot \hat{{\bm r}}_{ij}  {\bm \sigma}_j \cdot \hat{{\bm r}}_{ik} \right]  \nonumber \\
&& \times Z_\Lambda^\prime(r_{ij}) Z_\Lambda^\prime(r_{ik}),\label{eq:V3Ncoord}
\end{eqnarray}
where $S_{ij}$ and $ ( {\bf L}\cdot {\bm S})_{ij}$ are respectively the tensor and spin-orbit operators for particles $i$ and $j$, and the profile functions 
\begin{equation} \label{eq:cutoff}
Z_\Lambda(r)=\int \frac{d {\bf p}}{(2 \pi)^3} {\mathrm{e}}^{i {\bf p}\cdot {\bf r}} F({\bf p}^2;\Lambda),
\end{equation}
representing the cutoff in coordinate space, chosen as
\begin{equation}
  F({\bf p}^2,\Lambda)=\exp\left[-\left(\frac{{\bf p}^2}{\Lambda^2}\right)^2\right].
\end{equation}
In the following the value $\Lambda=500\,$MeV will be used.
The explicit expression for the N4LO LECs shift $\delta E_i$ of the three-body force can be found in Ref. \cite{Girlanda:2023znc} (see Eqs. (39)-(51)).
The induced contributions $\delta E_i$ are enhanced as compared to the genuine ones $E_i$, due to the presence of the nucleon mass factor, scaling as $m\sim O(\Lambda_\chi^2/p)$ (being $\Lambda_\chi$ hard or breakdown scale of the theory) in the Weinberg counting \cite{Weinberg:1991um}, which effectively promotes them to  N3LO.
In this work, the LECs $E_i$ will be thought of as constituted only of the induced contributions of the $P$-dependent $D_{16}$ and $D_{17}$.
Thus, at N3LO the $3N$ contact interaction depends on two combinations of the $2N$ LECs $D_i$, appearing in Eqs.~(\ref{eq:deltaD16})-(\ref{eq:deltaD17}), which cannot be determined from $2N$ scattering data, but have to be fitted to experimental observables in $A>2$ systems.

We investigate the sensitivity of polarization observables in low-energy $N-d$ scattering to the two $P$-dependent N3LO LECs. The AV18 potential is used as a representative $2N$ interaction, and the meaning of the LECs $C_S$ and $C_T$ in this framework is found treating the LO contact pionless  potential  as a very low-energy representation of the AV18 potential, with a local cutoff introduced. The values of $C_S$ and $C_T$ are thus taken from a fit  of the LO $2N$ contact interaction 
\begin{equation}
    \label{eq:vlambda}
    V^{(0)}_{2N, \Lambda}= \left[ C_S + C_T {\bm \sigma}_1 \cdot {\bm \sigma}_2 \right] Z_\Lambda (r) 
\end{equation}
to the singlet and triplet $n-p$ scattering lengths as predicted by the AV18 potential. In the above expression a local cutoff has been introduced as in Eq.~(\ref{eq:cutoff}).
From this procedure we get
\begin{equation}
    C_S=-66.53~{\mathrm{GeV}}^{-2}, \quad C_T=-3.47~{\mathrm{GeV}}^{-2}.
\end{equation}

The 3-body Schr$\mathrm{\ddot{o}}$dinger equation is solved as in  Ref.\cite{Girlanda:2023znc} employing the Hyperspherical Harmonic (HH) method. Below the deuteron breakup threshold, the $N-d$ scattering wave function is expressed as the sum of an internal $\Psi_C$ and an asymptotic part $\Psi_A$
\begin{equation}
\Psi_{LSJJ_z}=\Psi_C+\Psi_A\,,
\end{equation}
where the internal part $\Psi_C$ is expanded on the  HH basis as
\begin{equation}
\Psi_{C}=\sum_{\mu} c_{\mu} \Phi_{\mu}.\label{eq:psic}
\end{equation}
Here $\mu$ denotes all the quantum numbers required to define the basis element. 

The asymptotic part describes the relative motion between the nucleon and the deuteron at large distances, involving regular (R) and irregular (I) solutions.

Denoting these solution $\Omega^\lambda_LSJJz$, with $\lambda=R,I$ respectively, and defining
\begin{equation}
\Omega_{L S J  J_{z}}^{\pm}=i\Omega_{L S J  J_{z}}^{R} \pm \Omega_{L S J J_{z}}^{I},
\end{equation}
we have
\begin{equation} 
\Psi_{A}=\Omega_{L S J  J_{z}}^{-} +
\sum_{L^{\prime} S^{\prime}} \mathcal{S}_{L S, L^{\prime} S^{\prime}}^{J}(q) \Omega_{L^{\prime} S^{\prime} J J J_{z}}^{+}.\label{eq:PsiA}
\end{equation}
Here  $\mathcal{S}_{L S, L^{\prime} S^{\prime}}^{J}$  are the $S$-matrix elements and $q$ is defined as the modulus of the $N-d$ relative momentum. From the $S$-matrix it is possible to compute phase shifts and mixing angles, from which the scattering observables are calculated. The $S$-matrix in Eq.~(\ref{eq:PsiA}) and  the coefficients $c_{\mu}$ in Eq. (\ref{eq:psic}) are determined by the complex formulation of the Kohn variational principle \cite{KIEVSKY1997125}. This principle demands that a certain functional be stationary under variations of trial parameters, leading to a linear system whose solution provides the weights and coefficients.

The Hamiltonian is decomposed into $H_{2N}$ (kinetic energy plus AV18 $2N$ interaction with Coulomb potential) and $V_{3N,\Lambda}^{(2)}$ (containing $3N$ interaction induced by $D_{16}$ and $D_{17}$ terms). The linear system for coefficients involves the computation of matrix elements between HH basis elements and asymptotic functions.

A specific set of LECs allows the computation of the associated $S$-matrix for each $J^{\pi}$ state using the Kohn variational principle, providing observables at a specific energy.
\section{Fit results}\label{sec3}
The observables used in the fitting procedure include the $p-d$ differential cross section, the two vector analyzing powers $A_{y}$ and $i T_{11}$, the three tensor analyzing powers $T_{20}, T_{21}$, $T_{22}$ and the doublet and quartet  $n-d$ scattering lengths, with the experimental values $^2a_{nd}=(0.645\pm 0.003 \pm 0.007)$~fm \cite{scatteringlenght1} and $^4a_{nd} = (6.35 \pm 0.02)$~fm \cite{scatteringlenght2}. In particular we  fit the experimental doublet and quartet $n-d$ scattering lengths and the six $p-d$ scattering observables at center-of-mass energy $E_{\mathrm{cm}}=2\, \mathrm{MeV}$ \cite{shimizu}, amounting to $282$ experimental data.  In addition we also fix the $^3$H binding energy to 8.469~MeV. This value takes into account the contribution of the neutron-proton mass difference, which is not described in the HH method, amounting to $\sim 7$~keV, and additional amount of $\sim 6$~keV  from the truncation of the HH expansion.

In the case of the differential cross section, we introduce an overall normalization factor $Z$ in the definition of $\chi^{2}$. Specifically,

\begin{equation}
\chi^{2}=\sum_{i} \frac{\left(d_{i}^{\exp } / Z-d_{i}^{\mathrm{th}}\right)^{2}}{\left(\sigma_{i}^{\exp } / Z\right)^{2}},\label{eq:chi2}
\end{equation}
where $Z$ is determined from the minimization condition:
\begin{equation}
Z=\frac{\sum_{i} d_{i}^{\exp } d_{i}^{\mathrm{th}} /\left(\sigma_{i}^{\exp }\right)^{2}}{\sum_{i}\left(d_{i}^{\mathrm{th}}\right)^{2} /\left(\sigma_{i}^{\exp }\right)^{2}}.\label{eq:Z}
\end{equation}
Here, $d_{i}^{\exp/\mathrm{th}}$ represents the experimental data points and their theoretical predictions, respectively, while $\sigma_{i}^{\exp }$ is the experimental error. For other observables, we treat the normalization $Z=1.00 \pm 0.01$ as an additional experimental datum, considering the systematic uncertainty estimated as $1\%$ according to Ref.~\cite{shimizu}.

The fitting procedure involves a global 2-parameter fit including only  the $ P$-dependent $2N$ interaction. We also perform 3-parameter fits including the LO $3N$ contact interaction, 
\begin{equation} \label{eq:v3nlo}
    V^{(0)}_{3N,\Lambda}= E_0 \sum_{ijk} Z_\Lambda(r_{ij}) Z_\Lambda (r_{ik}),
\end{equation}
where the LEC $E_0$ introduces further flexibility to the fit and it is mainly determined by the $^3$H binding energy.

We start the iterative minimization procedure by solving the scattering problem for an initial random set of  $\alpha_4$ and $\alpha_5$ parameters, calculating the corresponding observables. Employing the POUNDerS algorithm \cite{pounders}, we repeat the process with different initial random $\alpha_i$ values, aiming to converge to the deepest minimum. 
The resulting $\chi^2$/d.o.f.  is $\sim 2.1 (1.9)$ for the two(three)-parameter fits. The fitted parameters $E_0$, $\alpha_4$ and  $\alpha_5$  can be read from the first column of Table \ref{tab:fits}.

\begin{table}
 \begin{tabular}{|c|c|c|}
\hline
Fitting procedure &  2-param. & 3-param.\\
\hline
$\chi^2$/d.o.f. & 2.1  & 1.9\\
\hline
$e_0$ & - &0.459\\
$\tilde \alpha_4 C_S$ & 1.751 & 1.894\\
$\tilde \alpha_5 C_S$ & -0.495  & -1.175\\
\hline
$^2a_{nd}$  (fm) &  0.573 & 0.599 \\
\hline
 \end{tabular}     
 \caption{Results of the 2-parameters and 3-parameters fits, the latter one obtained including the leading order $3N$ contact interaction. Here the fitted parameters $\alpha_i$ and  the corresponding values of the LO $3N$ contact LEC $E_0$  are dimensionless, i.e. $e_0 = E_0 F_\pi^4 \Lambda, \quad \tilde \alpha_i = \alpha_i F_\pi^4 \Lambda^3$ with $F_\pi=92.4$ MeV. In the last row  we report the value obtained for the  $n-d$ doublet scattering length $^2a_{nd}$, which should be compared with the experimental value $^2a_{nd}=(0.645\pm 0.003 \pm 0.007)$~fm \cite{scatteringlenght1}.} \label{tab:fits}
 \end{table}

\begin{figure}[h!]
    \centering
    \includegraphics[width=\linewidth]{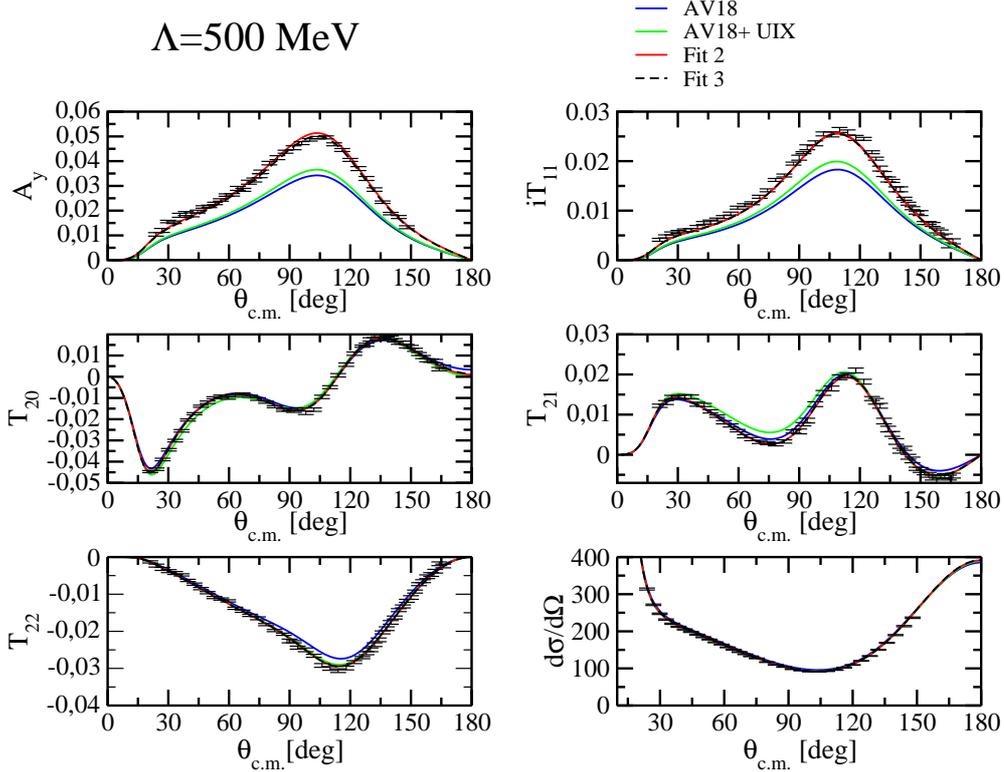}
    \caption{Proton and deuteron analyzing power  and differential cross-section at $E_{\mathrm{cm}}=2$~MeV. The red lines result from a global 2-parameter fit, the black lines represent the 3-parameters fit including the $E_0$ term, the blue lines are the predictions from the $2N$ AV18 potential, while the green lines are the predictions including also the $3N$ Urbana IX interaction. Experimental data are from Ref.~\cite{shimizu}. }
    \label{fig:fitnoce}
\end{figure}

Figure \ref{fig:fitnoce} shows the best fit curves for various analyzing powers and observables, compared to predictions from the $2N$ AV18 potential and the addition of the $3N$ Urbana IX interaction. The effective N3LO  $3N$ contact interaction induced by the $D_{16}$ and $D_{17}$ terms successfully addresses the $A_y$ problem, and the description of the vector analyzing power $i T_{11}$ is significantly improved.
We also conclude by saying that the introduction into the fit of the LEC $E_0$ corresponding to the three-body contact force at leading order does not substantially change the description of the experimental data, except for the observable $A_y$, as can be seen in Figure \ref{fig:fitnoce}.

\section{Conclusions}\label{sec4}

In this analysis, we have explored the  relativistic constraints on the $\mathcal{O}(p^4)$ $2N$ contact Lagrangian bringing out two $P$-dependent terms in the potential accompanied by two unconstrained LECs $D_{16}$ and $D_{17}$.
It should be emphasized that the above terms are not to be understood as relativistic corrections but for all intents are within the N3LO $2N$ contact potential.
These LECs, whose effect vanishes in the $2N$ center-of-mass frame, can play a crucial role in larger nuclear systems. The unitary equivalence to $3N$ contact operators implies a connection between these LECs and off-shell effects.
Using  an hybrid approach where the three-nucleon interaction, parametrized by $D_{16}$ and $D_{17}$ LECs, is considered alongside the phenomenological AV18 $2N$ potential, we fit experimental data on polarization observables in low-energy $p-d$ scattering, specifically focusing on the $A_y$ asymmetry.
The results indicate that the inclusion of terms represented by $D_{16}$ and $D_{17}$ in the three-nucleon interaction is crucial for accurately reproducing experimental data, in low-energy proton-deuteron scattering. This implies that these LECs can be  resolutive for the long-standing $A_y$ puzzle.
The $D_{16}$ and $D_{17}$ LECs on systems with $A > 2$ should be further quantified in future investigations. Of course it will be undoubtedly  intriguing to conduct a thorough reexamination of the aforementioned analysis within a fully consistent framework of ChEFT.






\bibliography{bibliography}

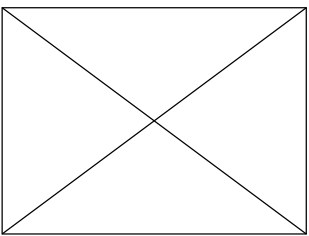
\end{document}